\documentstyle[a4,12pt,epsf]{article}

\newcommand{\nn}{\nonumber}

\newcommand{\be}{\begin{equation}}
\newcommand{\ee}{\end{equation}}
\newcommand{\bea}{\begin{eqnarray}}
\newcommand{\eea}{\end{eqnarray}}

\begin{document}
\begin{titlepage}
\renewcommand{\thefootnote}{\alph{footnote}}

\begin{flushright}
\parbox{40mm}{
KUNS-1903\\
KANAZAWA-04-06}
\end{flushright}

\begin{center}
\vspace*{10mm}
 
{\LARGE \bf Suppressed supersymmetry 
breaking terms in the Higgs sector}
\vspace{12mm}

{\large
Tatsuo~Kobayashi\footnote{E-mail address:
  kobayash@gauge.scphys.kyoto-u.ac.jp}
~and~~Haruhiko~Terao\footnote{E-mail address:
terao@hep.s.kanazawa-u.ac.jp}
}
\vspace{6mm}

{\it $^a$Department of Physics, Kyoto University,
Kyoto 606-8502, Japan}\\[1mm]

{\it $^b$Institute for Theoretical Physics, 
Kanazawa University,
Kanazawa 920-1192, Japan}

\vspace*{15mm}

\begin{abstract}
We study the little hierarchy between mass parameters 
in the Higgs sector and other SUSY breaking masses.
This type of spectrum can relieve the fine-tuning problem 
in the MSSM Higgs sector.
Our scenario can be realized by superconformal dynamics.
The spectrum in our scenario has significant implications 
in other phenomenological aspects like 
the relic abundance of the lightest neutralino and 
relaxation of the unbounded-from-below constraints.

\end{abstract}

\end{center}
\end{titlepage}
\pagestyle{plain}
\pagenumbering{arabic}
\setcounter{footnote}{0}

\section{Introduction}

Supersymmetric extensions of the Standard model 
is postulated as a solution for the hierarchy problem.
The radiative electroweak symmetry breaking is 
one of the attractive aspects of the supersymmetric model: 
the large top Yukawa coupling derives the Higgs 
mass to be tachyonic in the low energy \cite{Ibanez:fr}.
Within the framework of the minimal 
supersymmetric standard model (MSSM), 
the theoretical upper bound of the lightest 
Higgs boson mass $m_{h^0}$ at the tree-level 
is equal to the $Z$ boson mass $M_Z$.
On the other hand, the current experimental 
lower bound is 115 GeV.
A large correction to the Higgs mass can be 
obtained when the stop mass is as large as 500 GeV or more.
Such large stop mass squared induces the negative and 
comparable size of 
SUSY breaking Higgs scalar mass squared through the 
renormalization group (RG) effect.
That requires fine-tuning to lead to the weak scale from 
SUSY breaking parameters of  the magnitude of 500 GeV or 
higher mass scale.

So far, several mechanisms to solve the fine-tuning problem 
have been proposed \cite{Haber:1986gz,Polonsky:2000zt,casas,delgado,
fathiggs}.
One way is to assume 
additional quartic couplings of the Higgs fields other than 
$SU(2) \times U(1)_Y$ $D$-terms by extension of the gauge
sector, the Higgs sector and so on.
Another is to increase the theoretical prediction of 
the lightest Higgs scalar mass at tree level,
which may be achieved also by the additional quartic
couplings, and to make the radiative correction to Higgs
mass to be small.
In these mechanisms most soft scalar masses and 
$\mu$-parameter are often assumed to be 
relatively small to avoid the fine-tuning.

In this paper, we consider the little hierarchy  
between SUSY breaking parameters in the Higgs sector 
and others.
Then, we propose that such  
suppressed magnitude of mass parameters only 
in the Higgs sector, SUSY breaking Higgs masses
and also the $\mu$-parameter, 
improves the fine-tuning problem.
Such little hierarchy can be realized e.g. by 
superconformal dynamics, which must be decoupled 
around $O(1)-O(10)$ TeV.
Furthermore, the little hierarchy between SUSY breaking 
parameters lead to the spectrum different from 
the usual one, e.g. the minimal supergravity model.
Since the $\mu$-term is small, the lightest 
neutralino is not purely bino-like.
Also the constraint about charge/color breaking 
and unbounded-from-below directions can be ameliorated.

This paper is organized as follows.
In Section 2, we explain the fine-tuning problem 
in the MSSM, and we show our scenario to relax the 
problem.
We propose the little hierarchy between the mass parameters 
in the Higgs sectors and other SUSY breaking masses like 
the stop mass.
In section 3, we will show certain models to 
realize such little hierarchy by the superconformal 
dynamics.
In section 4, we will discuss generic phenomenological aspects 
of our scenario.
Section 5 is devoted to conclusion and discussions.

\section{Suppressed supersymmetry breaking 
in Higgs sector}

The Higgs sector of the MSSM has the following 
potential for the neutral components, 
$H^0_{u,d}$,
\begin{eqnarray}
V(H^0_u,H^0_d) &=& m_1^2 |H^0_d|^2 + m_2^2 |H^0_u|^2 -2 
\mu B  H^0_u H^0_d \nonumber \\ &+&
\frac{1}{8}(g^2+g'^2)(|H^0_d|^2 - |H^0_u|^2)^2,
\end{eqnarray}
with $m^2_1 = m^2_{H_d} + \mu^2$ and 
$m^2_2 = m^2_{H_u} + \mu^2$, where $m^2_{H_d}$, $m^2_{H_u}$ 
and $\mu B$ are SUSY breaking parameters, while 
$\mu$ is the supersymmetric mass parameter in 
the superpotential $\mu H^0_u H^0_d$.
The last term comes from the $SU(2)\times U(1)_Y$ 
$D$-terms.

If the following two conditions:
\begin{eqnarray}
m_1^2m_2^2 &<& (\mu B)^2, \\
m_1^2 + m_2^2 &>& 2|\mu B|,
\end{eqnarray}
are satisfied, the electroweak symmetry 
is broken.
The latter is the condition to avoid 
the direction of the potential unbounded from below along 
$\langle H^0_u \rangle = \langle H^0_d \rangle $.
Then the vacuum expectation value $v$, where 
$v^2 = \langle H^0_u \rangle^2 + \langle H^0_d \rangle^2$, 
is obtained by Higgs mass parameters as
\begin{equation}
v^2 = \frac{4}{g^2+g'^2} \left( -\mu^2 +
\frac{m_{H_d}^2 - m_{H_u}^2 \tan^2 \beta}{\tan^2 
\beta -1}\right),
\end{equation}
where $4/(g^2+g'^2) \approx 7$ at $M_Z$, and 
$\tan \beta \equiv \langle H^0_u \rangle 
/ \langle H^0_d \rangle$.
For a moderate value of $\tan \beta$, it reduces to 
\begin{equation}
v^2 \approx \frac{4}{g^2+g'^2} (-\mu^2 - m_{H_u}^2).
\label{weak-scale}
\end{equation}
Its magnitude $v$ itself should be derived as 
$v=174$ GeV.
The natural scale of $\mu^2$ and $m_{H_u}^2$ is 
$O(M_Z^2)$.

The theoretical prediction for the lightest CP-even Higgs 
boson mass $m_{h^0}$  at the tree level is equal to the 
$Z$ boson mass $M_Z$ or less.
The dominant one-loop correction is due to 
the top-stop mass splitting, because the top 
Yukawa is large.
Thus, the lightest Higgs mass 
$m_{h^0}$ at the one-loop is written as \cite{Haber:1990aw} 
\begin{equation}
m_{h^0}^2 \leq M_Z^2  + \frac{3}{4\pi^2}Y_t^4v^2
\ln \left(\frac{m^2_{\tilde t}}{m_t^2}\right),
\label{h-mass}
\end{equation}
where $Y_t$ is the top Yukawa coupling, and 
$m_t$ and $m_{\tilde t}$ are the top and stop masses, 
respectively.
The current experimental lower bound of $m_{h^0}$ is equal to 115 GeV.
That requires $m_{\tilde t} \geq 500$ GeV through 
Eq.~(\ref{h-mass}).

On the other hand, the SUSY breaking scalar mass 
$m_{H_u}$ receives the one-loop radiative 
correction between the high energy scale $\Lambda$ 
and the weak scale,
\begin{equation}
\Delta m_{H_u}^2 \sim -12 \frac{Y_t^2}{16\pi^2}
m_{\tilde t}^2 \ln \frac{\Lambda}{v}.
\label{log-h-mass}
\end{equation}
When we take e.g. $\Lambda = 10^{16}$ or $10^{18}$ GeV, 
we evaluate 
\begin{equation}
\Delta m_{H_u}^2 \sim - m_{\tilde t}^2.
\label{rad-h-mass}
\end{equation}
This rather large radiative correction is due to the 
long running scale, i.e. $\ln \frac{\Lambda}{v} \sim 30$.
Hence, the natural order of $|m_{H_u}^2|$ at the weak scale 
is equal to $O(m_{\tilde t}^2)$ when the initial value 
of $m_{H_u}^2$ at $\Lambda$ 
is comparable with $m_{\tilde t}^2$ or less.

Appearance of the same order soft scalar masses,
$m_{H_u}^2$ and $m_{\tilde t}^2$, at low energy
may be also explained in terms of 
the fixed point behavior.
The MSSM has the Pendleton-Ross fixed point \cite{Pendleton:as} for 
the top Yukawa coupling.
That is given as 
\begin{equation}
Y_t^2 =  (4\pi) \frac{7}{18}\alpha_3, 
\end{equation}
when we neglect the other gauge and Yukawa couplings.
At this fixed point, the SUSY breaking terms 
also satisfy the specific relation,
\begin{equation}
m^2_{\tilde t_L} + m^2_{\tilde t_R} +
m^2_{H_u} = M_3^2.
\end{equation}
Thus, around this fixed point, we have 
\begin{equation}
m^2_{H_u} \approx - m^2_{\tilde t_L} - m^2_{\tilde t_R} 
+ M^2_3.
\end{equation}
Note that the stop mass squared has 
the radiative correction $\Delta m_{\tilde t}^2$ 
between $\Lambda$ and the weak scale is of $O(M_3^2)$ 
e.g. for $\Lambda = 10^{16}-10^{18}$ GeV.
Again, the natural order of $|m_{H_u}^2|$ is 
of $O(m_{\tilde t}^2)$ or more.

Thus, the natural order of $|m_{H_u}^2|$ seems 
of $O(m_{\tilde t}^2)$ or more, and 
the lower mass bound of the lightest Higgs scalar 
field, $m_{h^0} \geq 115$ GeV, requires that 
$m_{\tilde t} \geq 500$ GeV.
That implies that the natural order of 
$|m_{H_u}^2|$ is of $O(500^2)$ (GeV)$^2$ or more.
However,  Eq.~(\ref{weak-scale}) requires 
that $\mu^2$ is also the same order 
and these two terms should be fine-tuned 
to lead to the weak scale.
Such a fine-tuning is often presented in terms of 
the following fine-tuning 
parameter \cite{Barbieri:1987fn,deCarlos:yy},
\begin{equation}
\Delta_{a^2} = \left|
\frac{a^2}{v^2} 
\frac{\partial v^2}{\partial a^2}\right| .
\end{equation}
For example, for $\mu^2, |m_{H_u}^2| \sim 500^2$ (GeV)$^2$, 
we have 
\begin{equation}
\Delta_{\mu^2}= O(100), \qquad \Delta_{m_{H_u}^2} = O(100) .
\end{equation}
Similarly, when  $m_{H_u}^2 \sim - m_{\tilde t}^2$, 
we have 
\begin{equation}
\Delta_{m_{\tilde t}^2}= O(100) .
\end{equation}
These values $\Delta_{a^2}$ increase linearly as 
$\mu^2, |m_{H_u}^2|$ and $m_{\tilde t}^2$ increase.

Thus, the combination of Eqs.~(\ref{weak-scale}), 
(\ref{h-mass}), (\ref{rad-h-mass}) as well as 
the experimental lower bound of the Higgs mass leads to 
unnaturalness.
This unnatural situation can be improved 
if at least one of these relations is modified.
Actually,  
several types of ideas have been proposed.
Some of them proposed additional quartic terms of 
the Higgs fields in the potential like 
$\delta \lambda_4 |H_u|^4$ \cite{Haber:1986gz,delgado,fathiggs}, 
e.g. by introducing singlet fields, which have couplings 
with $H_{u}H_d$ in the superpotential, 
or  by assuming new $D$-terms 
due to extra gauge symmetries, under which 
the Higgs fields have nontrivial 
representations.
Also, additional 
quartic terms can be generated by strong dynamics 
near the weak scale \cite{Polonsky:2000zt} or
low scale SUSY breaking \cite{casas}.
These additional quartic terms change 
the relation Eq.~(\ref{weak-scale}) and, therefore,
reduce the fine-tuning parameter directly.
In these cases, the relation Eq.~(\ref{h-mass}) is 
also changed, and the tree level lightest Higgs 
mass $m_{h^0}$ can be raised up
\footnote{Additional Higgs mass terms 
$\Delta m_{h_0}$ may be generated as hard SUSY breaking
terms through strong dynamics or low scale 
SUSY breaking.}.
Then, we do not need a large stop mass.
Thus relatively small scalar masses are allowed and
the fine-tuning can be ameliorated.

Here, we study the possibility for changing 
the radiative correction (\ref{rad-h-mass}) due to the stop mass, 
and we propose the scenario that SUSY breaking terms in 
the Higgs sector, e.g. $|m_{H_u}^2|$, are 
suppressed compared with other SUSY breaking terms
through some mechanism.
In the next section, we will show a certain 
model by the superconformal dynamics 
as an illustrating model to lead to suppressed 
SUSY breaking in the Higgs sector compared with 
other SUSY breaking terms.
We expect the scenario that $-m_{H_u}^2 \approx 100^2 - 200^2$ 
(GeV)$^2$ while sfermion masses are heavier, e.g. 
$m_{\tilde t} \geq 500$ GeV.
In this scenario, the $\mu$-parameter must 
also be suppressed, i.e.  $\mu^2 \sim -m_{H_u}^2$.
The other SUSY breaking parameters in the Higgs sector, 
$m_{H_d}$ and $B$, do not need to be the same order as 
$\mu$ and $O(m_{H_u})$.
Thus, we can expect the two cases: 
\begin{equation}
(a)~~~~ m_{\tilde t}^2, m_{H_d}^2 \gg 
\mu^2 \sim |m_{H_u}^2|,
\end{equation}
and
\begin{equation}
(b)~~~~ m_{\tilde t}^2 \gg 
m_{H_d}^2 \sim \mu^2 \sim |m_{H_u}^2|.
\end{equation}
Actually, the former case is realized in the next section 
by use of the superconformal dynamics.
The latter can also be realized by the superconformal 
dynamics, however, alternatively this may be 
realized e.g. in extra dimensional models, 
where the whole Higgs sector is separated away from the 
SUSY breaking source, while others like gaugino and matter fields 
are near the SUSY breaking source
\footnote{
The low energy effective theories given by the superconformal 
dyamics may be realized also in the warped geometry. 
The related setup has been studied in Ref.~\cite{Gherghetta:2003wm}.
}.
It is noted that $B^2 \gg \mu^2 \sim |m_{H_{u,d}}^2|$ 
does not lead to 
successful electroweak symmetry breaking, since
the Higgs potential becomes unbounded from below.
Therefore the initial value of $B$ must be properly 
suppressed in the latter case
\footnote{
Such a case is allowed if we have additional quartic couplings, 
e.g. $\delta \lambda_4 |H_u|^4$.
However the quartic coupling must be large enough to avoid
the fine-tuning problem.
}.

In the former case, the relation
$B^2 \gg \mu^2$, which is expected in the
scenario using superconformal dynamics, is allowed. 
The value of $\tan \beta$ is obtained as 
\begin{equation}
\frac{2 \tan \beta}{1+ \tan^2 \beta} = 
\frac{2\mu B}{m_1^2 + m_2^2} .
\end{equation}
In the case with $m_{H_d}^2, B^2 \gg \mu^2 
\sim |m_{H_u}^2|$, 
the value of $\tan \beta$ is obtained as 
\begin{equation}
\tan \beta = \frac{m^2_{H_d}}{\mu B}.
\end{equation}
Thus, we have a moderate or large value of 
$\tan \beta$, but not a
small value like $\tan \beta \sim 1$.
There are other phenomenological features in this scenario, 
and those will be discussed in Section 4 after showing 
an illustrating model in the next section.

Our primary purpose is to propose the scenario 
that the Higgs mass parameters are suppressed
at the low energy compared with other 
SUSY breaking parameters.
Such a scenario can be realized by the 
superconformal dynamics.
The superconformal sector must be decoupled 
around $O(1)-O(10)$ TeV.
This is because if the decoupling scale is 
higher, again we have a large radiative correction 
(\ref{rad-h-mass}).
Such a low decoupling scale of the superconformal 
sector may induce additional quartic coupling 
terms in the Higgs potential like 
$\delta \lambda_4 |H_u|^4$ as threshold corrections 
and/or additional 
Higgs mass corrections $\Delta m_{h_0}$.
In general these by-products can also work to 
relieve unnaturalness of the MSSM.

\section{Illustrating model}
Now we consider the case that the soft scalar mass
of the Higgs fields and also the $\mu$-parameter are suppressed  
due to strong dynamics, while other soft supersymmetry
breaking parameters are remained unsuppressed.
To be explicit, let us present certain models, in which
an extra superconformal gauge sector strongly coupled with 
the Higgs fields plays just this role. 
We introduce an $SU(4)$ gauge symmetry and 
the following extra matter fields other than the MSSM field 
content;
\be
\begin{array}{cclcccl}
T & : & ({\bf 4}, {\bf 3}, {\bf 1}, -1/3), & \hspace*{10mm} &
\bar{T} & : & (\bar{\bf 4}, \bar{\bf 3}, {\bf 1}, 1/3), \\
D & : & ({\bf 4}, {\bf 1}, {\bf 2}, 1/2), & \hspace*{10mm} &
\bar{D} & : & (\bar{\bf 4}, {\bf 1}, {\bf 2}, -1/2), \\
S & : & ({\bf 4}, {\bf 1}, {\bf 1}, 0), & \hspace*{10mm} &
\bar{S} & : & (\bar{\bf 4}, {\bf 1}, {\bf 1}, 0), \\
S' & : & ({\bf 4}, {\bf 1}, {\bf 1}, 0), & \hspace*{10mm} &
\bar{S}' & : & (\bar{\bf 4}, {\bf 1}, {\bf 1}, 0), \\
\end{array}
\ee
where the representations and charges under 
$SU(4) \times SU(3)_c \times SU(2)_W \times U(1)_Y$
are shown respectively in the brace. 
Unification of the MSSM gauge couplings in the perturbative regime
is not destroyed by these extra matter fields. 
\footnote{The running of the gauge couplings are affected slightly 
by large anomalous dimensions of the MSSM Higgs fields, though the 
energy scale where their couplings to the SC-sector become strong
is asuumed to be rather low. Therefore gauge coupling unification 
is not manifestly maintained and needs some further considerations.
} 
Also the model is vector-like and, therefore,
free from anomalies. 

For the moment, we consider this model with neglecting 
the MSSM gauge
and Yukawa interactions. Then the $SU(4)$ ($N_c=4$) gauge theory 
contains the flavor number $N_f=7$ of vector-like  matter 
fields with  (anti-)fundamental 
representation.
Therefore this theory belongs to the so-called superconformal
window, which is given by $(3/2)N_c < N_f < 3N_c$ for $SU(N_c)$
SQCD \cite{Seiberg:1994pq}.
The $SU(4)$ gauge coupling is attracted into a fixed point
at low energy and there the extra matter fields $\Phi = (T, D, S, S')$ 
acquire a large negative anomalous dimension 
$\gamma_{\Phi} = -5/7 < -1/2$.

We also assume that the extra matter fields carry odd matter
parity (R-parity) just like quarks and leptons. Then
allowed Yukawa couplings among extra matter fields
and the MSSM Higgs fields are strongly relevant because of 
the large anomalous dimensions of the extra matter fields.
It is well expected that there is
another infrared fixed point where the Higgs fields are
strongly coupled with the superconformal extra matter fields 
through Yukawa interactions.
Hereafter we assume that the Yukawa couplings become
large and the gauge theory approaches into
this fixed point around the energy scale $\Lambda_{\rm SC}$.
It is noted also that there are other relevant operators
than the Yukawa terms because of large anomalous dimensions of
the extra matter fields.
Actually it is a rather unclear matter what kinds of 
fixed point theories are possible.
Here we simply assume that the following superpotential is
realized at the infrared fixed point;
\bea
W &=& \lambda D \bar{S} H_u + \bar{\lambda} \bar{D} S H_d \nn \\
& & + \kappa(\bar{D} D)^2 + \kappa' (\bar{D}D)(\bar{S}S)
+ \kappa''(\bar{S} S)^2 \nn \\
& & + M_T \bar{T}T + M_D \bar{D}D + M_S \bar{S}S + M_{S'} \bar{S}'S'
+ \mu H_u H_d,
\eea
where the terms in the last line are supersymmetric mass terms.
Then the Higgs fields have large positive anomalous dimensions
due to the Yukawa couplings. In this case the anomalous dimensions
are fixed to be $\gamma_{H_u} = \gamma_{H_d} = 1$ and also
$\gamma_{D} = \gamma_{\bar{D}} = \gamma_{S} = \gamma_{\bar{S}} =-1/2$.
It should be noted here that $\mu$ is suppressed \cite{Kubo:2001cr}, 
while the supersymmetric masses $M_D$ and so on, which determine decoupling
scale of the superconformal sector from the MSSM sector,
are enhanced. The explicit relations may be written as 
\bea
\mu (\mu_{\rm IR}) &=& \left(
\frac{\mu_{\rm IR}}{\Lambda_{\rm SC}}
\right) \mu(\Lambda_{\rm SC}), \\
M_D (\mu_{\rm IR}) &=& \left(
\frac{\mu_{\rm IR}}{\Lambda_{\rm SC}}
\right)^{-1/2} M_D(\Lambda_{\rm SC}).
\eea
If the bare masses of the extra matter fields
are given to be the same order as bare $\mu$, 
the decoupling scale can be
much higher than the electroweak scale. 
For example, the Giudice-Masiero mechanism \cite{Giudice:1988yz} 
can induce the same order of supersymmetric masses, $\mu$ and 
$M_{T,D,S,S'}$ at the Planck scale, and 
furthermore, these are of the same order as 
those of SUSY breaking masses like gaugino and scalar masses.

It has been known that soft supersymmetry breaking parameters show
novel renormalization properties, when the theory stays at the
infrared attractive fixed point 
\cite{Karch:1998qa}-\cite{Terao:2001jw}. 
The gaugino mass and also the
A-parameters are suppressed with some powers of the renormalization
scale. The scalar masses enjoy sum rules at low 
energy irrespectively of their initial values.
In the case that matter fields $\phi_i$ of a superconformal gauge theory 
carries gauge representation $R_i$ and the superpotential contains
non-vanishing interactions of $\phi_i \phi_j \phi_k$ and so on,
the sum rules are found to be 
\cite{Kobayashi:2001kz,Nelson:2001mq,Terao:2001jw}
\be
\sum_i T(R_i) m^2_i \rightarrow 0, ~~~~~
m^2_i + m^2_j + m^2_k \rightarrow 0.
\ee
In the present model the sum rules show
that soft masses of $D, \bar{D}, S, \bar{S}, H_u, H_d$
are suppressed. Here we used the left-right symmetry and
assumed $m^2_D= m^2_{\bar{D}}$, $m^2_S= m^2_{\bar{S}}$.
In this model soft masses not only of $H_u$ but also of $H_d$
are suppressed.
The $\mu B $ term is suppressed like $\mu$, but 
the $B$ parameter itself is not suppressed by the superconformal 
dynamics.
Therefore the $B$-parameter must be small 
as an initial condition in order
to make successful electroweak symmetry breaking. 

It is possible to modify this  model so that the soft mass 
of $H_d$ is not suppressed. For this purpose let us introduce
a pair of mirror Higgs $H'_u$ and $H'_d$ and assign odd
matter parity to them.
Also we change the matter parity of the MSSM singlet fields
to $S:(+), \bar{S}:(-), S':(+), \bar{S}':(-)$.
Then the superpotential allowed by this parity assignment is
expected to be
\bea
W &=& \lambda D \bar{S} H_u + \lambda' \bar{D} S' H'_u 
 + \kappa(\bar{D} D)^2 + \kappa' (\bar{S}S')^2 \nn \\
& & + M_T \bar{T}T + M_D \bar{D}D + 
+ \mu H_u H_d + \mu' H'_u H'_d.
\eea
The Yukawa coupling with $H_d$ is forbidden by the matter
parity and therefore the soft scalar mass $m_{H_d}$ 
is not suppressed.
The mass $M_D$ as well as $M_T$ is enhanced and 
the superconformal sector decouples from the MSSM sector
and enters confinement phase below this scale.

As well as the former model, the $\mu$-parameter is suppressed.
However the factor becomes milder and it is given at the
decoupling scale as
\be
\mu (M_D) = \left(
\frac{M_D}{\Lambda_{\rm SC}}
\right)^{1/2} \mu(\Lambda_{\rm SC}), 
\ee
since only $H_u$ has a large anomalous dimension.
Explicitly we may suppose that the decoupling scale $M_D$
is around a few TeV and $\Lambda_{\rm SC}$ is around 
$10 \sim 10^2$ TeV. Then the $\mu$-parameter is suppressed
about one order.

However what we are most concerned with is the MSSM
radiative corrections to the low energy $m^2_{H_u}$ and
their dependence on $m^2_{\tilde{t}}$.
When the MSSM interactions are switched on, the soft scalar
mass squared of $H_u$, $m^2_{H_u}$, is not simply suppressed.
The relevant part of the RG flow equation for $m^2_{H_u}$
will be given by
\be
\mu \frac{d m^2_{H_u}}{d \mu} 
= \Gamma_{H_u} m^2_{H_u} + 
\frac{12}{16\pi^2} Y^2_t m^2_{\tilde{t}}.
\label{floweq}
\ee
Here the factor $\Gamma_{H_u}$ indicates the degree of
power suppression of the scalar mass by the superconformal 
dynamics. 
(See Ref.\cite{Kobayashi:2001kz,Kobayashi:2002iz,Terao:2001jw}.)
The non-perturbative dynamics makes it difficult to know
this factor, but it would be similar to the anomalous 
dimension $\gamma_{H_u}$.
Let us evaluate $m^2_{H_u}$ at the decoupling scale $M_D$ 
by using Eq.~(\ref{floweq}). If we ignore renormalization
of $Y_t$ and $m^2_{\tilde{t}}$, then
it is immediately seen that the soft scalar mass
behaves as\footnote{Similar behavior has been studied in 
Ref.\cite{Kobayashi:2001kz,Kobayashi:2002iz,Terao:2001jw}.}
\be
m^2_{H_u}(M_D) \rightarrow 
- \frac{12}{16\pi^2 \Gamma_{H_u}}Y^2_t m^2_{\tilde{t}}(M_D).
\ee
In practice, the top Yukawa coupling is also suppressed 
just like the $\mu$-parameter. 
However it may be shown that $m^2_{H_u}(M_D)$
is given similarly even if taking account of the power correction
to the Yukawa coupling.
This equation should be compared with Eq.~(\ref{log-h-mass}).
Then it is noted that the large logarithmic factor disappears.
This is because that the large radiative correction 
at high energy is erased by the attractive nature of
superconformal dynamics,
and the low energy Higgs mass is determined at the
decoupling scale.
Here we would stress again that the radiative correction
to $m^2_{H_u}$ largely enhanced by the huge scale difference
is the biggest factor causing the MSSM fine
tuning problem. 
Therefore the degree of fine tuning with respect to 
$m^2_{\tilde{t}}$ (or gluino mass) can be remarkably improved
by the superconformal dynamics. 
Also fine-tuning with respect 
to $\mu$-parameter is dissolved simultaneously.

We may expect that large corrections to the Higgs mass and also
to the Higgs quartic coupling are induced through decoupling
of the superconformal sector, since $H_u$ is coupled very
strongly with this sector
\footnote{The effective Higgs quartic coupling generated by decoupling
of extra gauged matter fields has been discussed in Ref.~\cite{delgado}.
However the soft supersymmetry breaking masses of the extra matter fields
need to be large as 10 TeV order in order to ameliorate the 
fine tuning problem.
}.
However note that the superconformal gauged matter fields  coupled to
$H_u$ ($D, S$ in the explicit model) undergo suppression of the
scalar masses. 
According to the naive dimensional analysis \cite{nda},
the correction to the Higgs soft mass due to decoupling
is roughly evaluated as 
$\delta m^2_{H_u} \sim m^2_{H_u} m^2_{D}/M^2_D$,
which is much smaller than $m^2_{H_u}$.
Therefore suppression of the soft mass of Higgs field to the
electroweak scale is not destroyed by the decoupling
effect of the superconformal sector.
Similarly, however, the correction to the Higgs quartic coupling
would be small in these models.

\section{Generic aspects}

In the previous section, we have shown concrete models 
leading to suppressed Higgs soft scalar masses and the 
suppressed $\mu$-term, compared with other SUSY breaking masses.
Here we collect results for the case that only the Higgs fields 
couple with generic superconformal sector, but 
MSSM matter fields do not couple.
Then, we will discuss some phenomenological aspects of our 
scenario.
The $\mu$ term is suppressed as 
\begin{equation}
\mu (M_D) \approx \left(
\frac{M_D}{\Lambda_{\rm SC}}
\right)^{\gamma_{H_u}+ \gamma_{H_d}} \mu(\Lambda_{\rm SC}), 
\end{equation}
where $\gamma_{H_u}$ and $\gamma_{H_d}$ are anomalous dimensions of 
$H_{u,d}$ induced by the superconformal sector.
If $H_d$ has negligible couplings, we have $\gamma_{H_d} \sim 0$.
The soft scalar mass of $H_u$ is also suppressed at $M_D$ as 
\begin{equation}
m^2_{H_u}(M_D) \sim 
- \frac{12}{16\pi^2 \Gamma_{H_u}}Y^2_t m^2_{\tilde{t}}(M_D),
\end{equation}
independently of initial values.
The soft scalar mass of $H_d$ is also suppressed when $H_d$ 
couples with the superconformal sector
\footnote{
The RG flow of the top Yukawa coupling is the similar to 
one of $\mu$.
We need $Y_t (v) \sim 1$ to realize the top mass.
That implies that $Y_t (\Lambda_{\rm SC})$ must be large,  
depending 
how much hierarchy we expect between $\mu$ and other 
SUSY breaking masses e.g. $m_{\tilde t}$.
}.
When $Y^2_t/\Gamma_{H_u} \sim 1$ at $M_D$, 
we evaluate 
\begin{equation}
m^2_{H_u}(M_D) \sim 
- 0.08 \times m^2_{\tilde{t}}(M_D).
\end{equation}
It is expected that $\mu(\Lambda_{\rm SC}) \sim 
m_{\tilde t}$ by a certain mechanism like the 
Giudice-Masiero mechanism.
Then, the little hierarchy among $m_{H_u}^2$, $\mu^2$ and 
other SUSY breaking terms like $m_{\tilde t}^2$ can be realized.
For example, we would have $-m^2_{H_u}(M_D) \sim (100)^2$ (GeV)$^2$
for $m_{\tilde{t}}(M_D) \sim 500$ GeV.
Compared with Eq.~(\ref{log-h-mass}), 
the fine-tuning parameter 
$\Delta_{m_{\tilde t}^2(M_D)}$ 
in our present scenario is 
reduced by the large logarithmic factor $\ln (\Lambda/v) \sim 30$.
For example, we have $\Delta_{m_{\tilde t}^2(M_D)}  \leq O(10)$ 
for $m_{\tilde t} \sim 500$ GeV. 
However, the decoupling scale $M_D$ should not be far away from 
the weak scale.

In this section, we discuss other phenomenological 
implications of our spectrum.
In the minimal supergravity model, 
most of parameter regions lead to a large value of 
$\mu$ compared with the bino mass.
That implies that the lightest neutralino is 
purely bino-like, which
is the lightest superpartner (LSP).
In contrast, our scenario leads to a small value of 
$\mu$.
Thus, the lightest neutralino can be 
a mixture of the higgsino and bino including the case 
with the purely higgsino.
Furthermore, the next lighter neutralino and 
the lightest chargino can have similar masses.
This brings a drastic change in the relic abundance 
of the LSP \cite{Mizuta:1992qp}.
However the mass spectrum depends on the detail of models,  
specially on the gaugino masses. Explicit
considerations on the relic abundance is beyond our
scope.

Also the situation on the charge and/or color breaking 
and unbounded-from-below constraints can change.
The most serious unbounded-from-below direction 
of the MSSM scalar potential involves the Higgs and 
slepton fields $\{H_u,\tilde \nu_{L_i},\tilde e_{L_j}, 
\tilde e_{R_j}\}$ , i.e. the so-called UFB-3 
direction \cite{Casas:1995pd,Casas:1996wj}.
We use the stationary conditions 
$\partial V/\partial  \phi =0$ for $\phi = \{\tilde \nu_{L_i},\tilde e_{L_j}, 
\tilde e_{R_j}\}$, and write the field values of 
$\{\tilde \nu_{L_i},\tilde e_{L_j}, \tilde e_{R_j}\}$  
in terms of $H_u$.
Then, the potential along the UFB-3 direction is written as 
\begin{equation}
V_{\rm UBF-3} = (m_{H_u}^2 + m_{\tilde L_i}^2)|H_u|^2 
+ \frac{|\mu|}{Y_{e_j}}(m_{\tilde L_j}^2 +m_{\tilde e_j}^2 + 
m_{\tilde L_i}^2 )|H_u| - \frac{2 m_{\tilde L_i}^4 }{g^2 + g'^2} ,
\end{equation}
if the following inequality is satisfied,
\begin{equation}
|H_u| > \sqrt{\frac{\mu^2}{4Y_{e_j}^2} + 
\frac{4m_{\tilde L_i}^2 }{g^2 + g'^2}} - \frac{|\mu|}{2Y_{e_j}} ,
\end{equation}
where $Y_{e_j}$ is the lepton sector Yukawa coupling.
Otherwise, 
\begin{equation}
V_{\rm UBF-3} = m_{H_u}^2 |H_u|^2 
+ \frac{|\mu|}{Y_{e_j}}(m_{\tilde L_j}^2 +m_{\tilde e_j}^2 )|H_u| 
+\frac{1}{8}(g^2 + g'^2)\left[ |H_u|^2 + \frac{|\mu|}{Y_{e_j}} 
|H_u| \right]^2 .
\end{equation}

The Higgs soft mass squared $m^2_{H_u}$
is strongly driven negatively 
by the stop mass, that is, the gluino mass, 
while radiative corrections to $m_{\tilde L_j}^2$ due to 
the wino mass is small.
Therefore the quantity 
$m^2_{H_u} + m_{\tilde L_i}^2$ tends to be negative.
Thus, the large negative contribution due to $m_{H_u}^2$ 
is serious along the UFB-3 direction, 
if the initial condition of $m_{H_u}^2$ and $m_{\tilde L_j}^2$ 
are of the same order like the minimal 
supergravity model \cite{Casas:1996wj}.
In our scenario, we have the spectrum 
$|m_{H_u}^2| < m_{\tilde L_j}^2$ naturally.
Therefore the UFB-3 bound can be relaxed also.

\section{Conclusion and discussions}
The experimental lower bound of the lightest Higgs
boson leads that the mass parameters in the
Higgs sector of the MSSM must be fine-tuned to a few 
\%-order.
There have been various modifications to avoid this 
problem so far. In the most cases, additional
contributions for the quartic couplings of the
Higgs bosons are introduced and the relatively
light soft scalar masses are also assumed.

Here we have studied the scenario that only 
the $\mu$ term and 
SUSY breaking mass terms in the Higgs sector 
are suppressed 
compared with other SUSY breaking masses, e.g. the stop mass.
Our scenario can be realized by the superconformal dynamics,
which can alter the RG running for soft scalar mass of the
Higgs fields at low energy.
The large radiative correction to the Higgs mass 
generated at higher energy can be erased.
Its decoupling scale must be a few TeV or less.
Our illustrating models can naturally generate such 
hierarchy, $\mu^2, m_{H_u}^2 \ll m_{\tilde t}^2 \ll M_D^2$,
in the case that all of SUSY breaking masses have the same 
order of 
initial values and the $\mu$ term and supersymmetric 
mass terms in the superconformal sector are induced, e.g. by 
the Giudice-Masiero mechanism.
Thus the mass spectrum in our scenario can relieve the 
fine-tuning problem.
In general threshold effects due to the decoupling may 
induce additional quartic couplings of the Higgs 
fields and affect the lightest Higgs mass.
These effects altogether work to relieve the fine-tuning 
problem, although the decoupling effects seem to be 
small in the models explicitly given here.

As other phenomenological aspects, small values of 
$\mu^2$ and $m_{H_u}^2$ are significant.
The smallness of $\mu$ affects on  mixing of gaugino and higgsino 
in the neutralino and chargino states.
That would drastically change the relic abundance of the LSP.
Furthermore, compared with slepton masses
the smallness of $|m_{H_u}^2|$ is favorable from the viewpoint of 
the most serious unbounded-from-below direction.

\section*{Acknowledgment}

The authors thank D.~Suematsu and M.~M.~Nojiri for discussions
and comments.
T.~K.\/ is supported in part by the Grant-in-Aid for 
Scientific Research  (\#14540256) and 
the Grant-in-Aid for 
the 21st Century COE ``The Center for Diversity and 
Universality in Physics'' from Ministry of Education, Science, 
Sports and Culture of Japan.
H.~T.\/ is supported in part by the Grant-in-Aid for 
Scientific Research  (\#13135210)
from Ministry of Education, Science, 
Sports and Culture of Japan.


\begin{thebibliography}{99}


\bibitem{Ibanez:fr}
L.~E.~Ibanez and G.~G.~Ross,
Phys.\ Lett.\ B {\bf 110}, 215 (1982);
L.~Alvarez-Gaume, M.~Claudson and M.~B.~Wise,
Nucl.\ Phys.\ B {\bf 207}, 96 (1982);
K.~Inoue, A.~Kakuto, H.~Komatsu and S.~Takeshita,
Prog.\ Theor.\ Phys.\  {\bf 67}, 1889 (1982);
Prog.\ Theor.\ Phys.\  {\bf 68}, 927 (1982)
[Erratum-ibid.\  {\bf 70}, 330 (1983)];
Prog.\ Theor.\ Phys.\  {\bf 71}, 413 (1984).


\bibitem{Haber:1986gz}
See e.g., 
H.~E.~Haber and M.~Sher,
Phys.\ Rev.\ D {\bf 35}, 2206 (1987);
M.~Drees,
Phys.\ Rev.\ D {\bf 35} (1987) 2910;
K.~S.~Babu, X.~G.~He and E.~Ma,
Phys.\ Rev.\ D {\bf 36}, 878 (1987);
J.~R.~Espinosa and M.~Quiros,
Phys.\ Lett.\ B {\bf 279}, 92 (1992);
Phys.\ Lett.\ B {\bf 302}, 51 (1993);
Phys.\ Rev.\ Lett.\  {\bf 81}, 516 (1998);
M.~Cvetic, D.~A.~Demir, J.~R.~Espinosa, L.~L.~Everett and P.~Langacker,
Phys.\ Rev.\ D {\bf 56}, 2861 (1997)
[Erratum-ibid.\ D {\bf 58}, 119905 (1998)];
P.~Binetruy and C.~A.~Savoy,
Phys.\ Rev.\ D {\bf 277}, 453 (1992);
G.~L.~Kane, C.~F.~Kolda and J.~D.~Wells,
Phys.\ Rev.\ Lett. {\bf 70}, 2686 (1993).

\bibitem{Polonsky:2000zt}
N.~Polonsky and S.~f.~Su,
Phys.\ Rev.\ D {\bf 63}, 035007 (2001);
Phys.\ Lett.\ B {\bf 508}, 103 (2001).



\bibitem{casas}
A.~Brignole, J.~A.~Casas, J.~R.~Espinosa and I.~Navarro,
Nucl.\ Phys.\ B {\bf 666}, 105 (2003) ;
J.~A.~Casas, J.~R.~Espinosa and I.~Hidalgo,
JHEP {\bf 0401} (2004) 008.

\bibitem{delgado}
P.~Batra, A.~Delgado, D.~E.~Kaplan and T.~M.~P.~Tait,
JHEP {\bf 0402} (2004) 043.

\bibitem{fathiggs}
R.~Harnik, G.~D.~Kribs, D.~T.~Larson and H.~Murayama,
hep-ph/0311349.


\bibitem{Haber:1990aw}
H.~E.~Haber and R.~Hempfling,
Phys.\ Rev.\ Lett.\  {\bf 66}, 1815 (1991);
Y.~Okada, M.~Yamaguchi and T.~Yanagida,
Prog.\ Theor.\ Phys.\  {\bf 85}, 1 (1991);
Phys.\ Lett.\ B {\bf 262}, 54 (1991).
J.~R.~Ellis, G.~Ridolfi and F.~Zwirner,
Phys.\ Lett.\ B {\bf 257}, 83 (1991);
Phys.\ Lett.\ B {\bf 262}, 477 (1991).


\bibitem{Pendleton:as}
B.~Pendleton and G.~G.~Ross,
Phys.\ Lett.\ B {\bf 98}, 291 (1981).


\bibitem{Barbieri:1987fn}
R.~Barbieri and G.~F.~Giudice,
Nucl.\ Phys.\ B {\bf 306}, 63 (1988).



\bibitem{deCarlos:yy}
Also see, e.g.,
B.~de Carlos and J.~A.~Casas,
Phys.\ Lett.\ B {\bf 309}, 320 (1993)
M.~Olechowski and S.~Pokorski,
Nucl.\ Phys.\ B {\bf 404}, 590 (1993);
G.~W.~Anderson and D.~J.~Castano,
Phys.\ Lett.\ B {\bf 347}, 300 (1995);
G.~W.~Anderson and D.~J.~Castano,
Phys.\ Rev.\ D {\bf 52}, 1693 (1995) ;
P.~H.~Chankowski, J.~R.~Ellis and S.~Pokorski,
Phys.\ Lett.\ B {\bf 423}, 327 (1998);
R.~Barbieri and A.~Strumia,
Phys.\ Lett.\ B {\bf 433}, 63 (1998);
S.~Dimopoulos and G.~F.~Giudice,
Phys.\ Lett.\ B {\bf 357}, 573 (1995);
G.~L.~Kane and S.~F.~King,
Phys.\ Lett.\ B {\bf 451}, 113 (1999).

\bibitem{Gherghetta:2003wm}
T.~Gherghetta and A.~Pomarol,
Phys.\ Rev.\ D {\bf 67}, 085018 (2003).


\bibitem{Seiberg:1994pq}
N.~Seiberg,
Nucl.\ Phys.\ B {\bf 435}, 129 (1995);
K.~A.~Intriligator and N.~Seiberg,
Nucl.\ Phys.\ Proc.\ Suppl.\  {\bf 45BC}, 1 (1996).

\bibitem{Kubo:2001cr}
J.~Kubo and D.~Suematsu,
Phys.\ Rev.\ D {\bf 64}, 115014 (2001).


\bibitem{Giudice:1988yz}
G.~F.~Giudice and A.~Masiero,
Phys.\ Lett.\ B {\bf 206}, 480 (1988).


\bibitem{Karch:1998qa}
A.~Karch, T.~Kobayashi, J.~Kubo and G.~Zoupanos,
Phys.\ Lett.\ B {\bf 441}, 235 (1998).

\bibitem{Luty:1999qc}
M.~A.~Luty and R.~Rattazzi,
JHEP {\bf 9911}, 001 (1999).



\bibitem{Nelson:2000sn}
A.~E.~Nelson and M.~J.~Strassler,
JHEP {\bf 0009}, 030 (2000).

\bibitem{Kobayashi:2001kz}
T.~Kobayashi and H.~Terao,
Phys.\ Rev.\ D {\bf 64}, 075003 (2001).


\bibitem{Nelson:2001mq}
A.~E.~Nelson and M.~J.~Strassler,
JHEP {\bf 0207}, 021 (2002).

\bibitem{Kobayashi:2002iz}
T.~Kobayashi, H.~Nakano, T.~Noguchi and H.~Terao,
Phys.\ Rev.\ D {\bf 66}, 095011 (2002).


\bibitem{Terao:2001jw}
H.~Terao,
arXiv:hep-ph/0112021.




\bibitem{nda}
M.~A.~Luty, Phys.~Rev. {\bf D57} (1998) 1531;
A.~G.~Cohen, D.~B.~Kaplan and A.~E.~Nelson,
Phys.~Lett.{\bf B412} (1997) 301.



\bibitem{Mizuta:1992qp}
S.~Mizuta and M.~Yamaguchi,
Phys.\ Lett.\ B {\bf 298}, 120 (1993) ;
J.~Edsjo and P.~Gondolo,
Phys.\ Rev.\ D {\bf 56}, 1879 (1997).



\bibitem{Casas:1995pd}
J.~A.~Casas, A.~Lleyda and C.~Munoz,
Nucl.\ Phys.\ B {\bf 471}, 3 (1996).

\bibitem{Casas:1996wj}
J.~A.~Casas, A.~Lleyda and C.~Munoz,
Phys.\ Lett.\ B {\bf 380}, 59 (1996);
Phys.\ Lett.\ B {\bf 389}, 305 (1996).








\end{thebibliography}
\end{document}